\documentclass[raggedfooter,pre,aps,twocolumn,superscriptaddress,noshowpacs,showkeys,floatfix,nofootinbib]{revtex4}
\usepackage{amsbsy}
\usepackage{amsmath} 
\usepackage{amssymb}
\usepackage{array}
\usepackage{bm}
\usepackage{algorithm}
\usepackage{algorithmic}
\usepackage{slashbox}
\usepackage[dvips]{graphicx}

\renewcommand{\algorithmicrequire}{\textbf{Input:}}
\renewcommand{\algorithmicensure}{\textbf{Output:}}

\DeclareMathAlphabet{\mathpzc}{OT1}{pzc}{m}{it}

\newcounter{todos}

\newcommand*{\numtodos}{0}
\makeatletter
\AtBeginDocument{%
 \AtEndDocument{%
  \immediate\write\@mainaux{%
   \string\gdef\string\numtodos{\number\value{todos}}%
  }%
 }%
}

\begin{document}
\title{Exact Ground States of \protect{\\} Large Two-Dimensional Planar Ising Spin Glasses}

\author{G. Pardella}
\author{F. Liers}
\affiliation{Institut f{\"u}r Informatik, Universit{\"a}t zu K{\"o}ln,
Pohligstra{\ss}e 1, D-50969 K{\"o}ln,Germany.}
\email[E-mail: ]{{pardella|liers}@informatik.uni-koeln.de}
\homepage[Visit: ]{http://cophy.informatik.uni-koeln.de/}

\date{\today}

%---------------------------------------------------------------------------%
%---------------------------------------------------------------------------%

\begin{abstract}
  Studying spin-glass physics through
  analyzing their ground-state properties has a long history.
  Although there exist polynomial-time algorithms for the
  two-dimensional planar case, where the problem of finding ground
  states is transformed to a minimum-weight perfect matching
  problem, the reachable system sizes have been limited both by the
  needed CPU time and by memory requirements. In this work, we present an
  algorithm for the calculation of exact ground states 
  for two-dimensional Ising spin glasses with free boundary conditions
  in at least one direction. The algorithmic foundations of the
  method date back to the work of Kasteleyn from the 1960s for 
  computing the complete partition function of the Ising model. Using
  Kasteleyn cities, we 
  calculate exact ground states for huge two-dimensional planar
  Ising spin-glass lattices (up to $3000^2$ spins) within reasonable
  time. According to our knowledge, these are the largest sizes
  currently available.
  Kasteleyn cities were recently
  also used by Thomas and Middleton in the context of extended ground
  states on the torus. Moreover, they show that the method can
  also be used for computing ground states of planar
  graphs. Furthermore, we point out that the correctness of
  heuristically computed ground states can easily be
  verified. Finally, we evaluate the solution quality of heuristic
  variants of the Bieche et al. approach.
\end{abstract}

%---------------------------------------------------------------------------%
%---------------------------------------------------------------------------%


%---------------------------------------------------------------------------%
%---------------------------------------------------------------------------%

\maketitle

%---------------------------------------------------------------------------%
%---------------------------------------------------------------------------%

\section{Introduction}

The determination of spin-glass ground states has raised the interest
of both physicists and computer scientists. For an introduction
we refer to [\onlinecite{hartmannrieger02,rieger98, hartmann08}]. On the one hand, an
analysis of the ground-state properties sheds light on the ruling
physics of the system. On the other hand, several different algorithms
have been developed and used for the ground-state determination of
different models. 

For the two-dimensional Edwards-Anderson model [\onlinecite{EA75}] (EA) 
with free boundaries in at least one direction, ground states can be 
determined exactly with fast algorithms. In fact, the problem is solvable in 
time bounded by a polynomial in the size of the input. The latter can be achieved by a
transformation to a well known graph theoretical problem \textemdash the
minimum-weight matching problem, for which efficient implementations
exist. For general non-planar or three- or higher-dimensional lattices,
however, calculating exact ground states is \textsc{NP}-hard [\onlinecite{papadimitriou82}]. 
Loosely speaking, this means we cannot expect to be able to design a polynomial-time
solution algorithm. In practice, one can use, e.g. branch-and-cut algorithms [\onlinecite{ljrr04}].

In this work, we focus on the polynomially solvable case of two-dimensional
lattices with free boundaries in at least one direction. We first
review and compare the main known approaches which are those of Bieche et
al. [\onlinecite{bieche80}] and of Barahona 
[\onlinecite{barahona82a, barahona82b}]. Then we present the approach inspired 
by Kasteleyn [\onlinecite{kasteleyn63}]. All method basically follow the same idea: 
An associated graph is constructed in which a minimum-weight perfect matching is
determined that is used to construct an exact ground
state. Differences occur in the constructed associated graph.
It turns out that the approach inspired by Kasteleyn is the most favorable.
In fact, using the latter method, we can determine exact ground states 
for lattice sizes up to $3000^2$, while the possible sizes computed earlier 
with heuristic variants of the approach of Bieche et al. were considerably smaller. 
In a forthcoming article [\onlinecite{gregorfraukeflorent08}], 
we will analyze the physics of the system.
Kasteleyn cities were recently also used by Thomas and Middleton
[\onlinecite{middleton07}]. While focussing on extended ground states
on the torus, they show that the Kasteleyn-city approach can be
successfully used in the planar case, too.  Furthermore, they compared their
implementation with an implementation of Barahona's method. It turned
out that the approach with Kasteleyn cities is less memory consumptive
and faster. Apart from this recent work, we are not aware of other
computational studies using Barahona's method.

We show how to either prove correctness of heuristically determined
ground states or how to correct them using linear programming. Despite
the fact that this is fast, it is still advantageous to use the method
based on Kasteleyn cities. Finally, we evaluate the quality of the
solutions generated with heuristic variants of the Bieche et al.
approach.

The outline of the article is as follows. In Section \ref{sec:model}, we
introduce the model. In Section \ref{sec:prelim} we introduce
definitions necessary for the literature review in Section
\ref{sec:algorithms}. Finally, we report the results in Section
\ref{sec:compResults}.

%---------------------------------------------------------------------------%
%---------------------------------------------------------------------------%
\section{The Model}\label{sec:model}
In the Edwards-Anderson model $N$ spins are placed
on a lattice. We focus on quadratic ($N=L^2$) lattices with free
boundary conditions in at least one direction. Toric boundary
conditions may be applied to at most one lattice axis. The Hamiltonian
of the system is
\begin{equation}
\mathcal{H} = -\sum_{<i,j>} J_{ij}S_{i}S_{j},
\end{equation}
where the sum runs over all nearest-neighbor sites. Each spin $S_i$ is
a dynamical variable which has two allowed states, $+1$ and $-1$. The
coupling strengths $J_{ij}$ between spins $i \text{ and } j$ are
independent identically distributed random variables following some
probability distribution. The concentration of anti-ferromagnetic
($J_{ij} < 0$) and ferromagnetic bonds ($J_{ij} > 0$) depends on the
underlying distribution. The Gaussian and the bimodal $\pm J$
distributions are often used. 
A spin configuration attaining the global minimum of the energy
function $\mathcal{H}$ is called a ground state. 

%---------------------------------------------------------------------------%
%---------------------------------------------------------------------------%
\section{Preliminaries}\label{sec:prelim}

In this section, we briefly summarize some basic definitions from graph
theory. For further details, we refer to [\onlinecite{harary69,
  diestel06, bollobas98}] and the references therein.
We associate a spin-glass instance with a \textit{graph} $G=(V,E)$
with vertices $V$ (spin sites) and edges $E$ (bonds). The edge set
consists of unordered pairs $(i,j)$, with $i,j\in V$. More
specifically, for two-dimensional $K\times L$ lattices with free
boundaries, the graph is called a \textit{grid graph} and is denoted by
$G_{K,L}$. 
In case periodic boundaries are present in one
direction, we call the graph a \textit{half-torus} or a \textit{ring}. The
\textit{degree} $\textit{deg}(v)$ of a vertex $v$ is the number of edges
$(v,w_i) \in E$ incident at $v$. A \textit{path}, $\pi =
v_1,v_2,...,v_k,\ v_i \in V$, is a sequence of vertices such that
$(v_1,v_2), (v_2,v_3), ..., (v_{k-1},v_k)$ are edges of $G$ and the
$v_i$ are distinct. A closed path $\pi = v_1,v_2,...,v_k,v_1$ is a
\textit{cycle}.

In many applications a rational cost or a weight $w(e)$ is associated with
an edge $e$. Let $G=(V,E)$ be a weighted graph.
For each (possibly empty) subset $Q\subseteq V$, a \textit{cut}
$\delta(Q)$ in $G$ is the set of all edges with one vertex in $Q$
and the other in $V\setminus Q$.  The weight of a cut is given by
$w(\delta(Q)) = \sum\limits_{e=(v,w) \in \delta(Q)} w(e)$. 
A \textit{minimum cut} (\textsc{min-cut}) asks for a cut $\delta(Q)$ with minimum
weight among all vertex sets $Q\subseteq V$. Let $K_n$ denote the \textit{complete
  graph} with $n$ vertices and edge set $E = V \times V$.  A \textit{ 
  subgraph} of $G$ is a graph $G_H$ such that every vertex of
$G_H$ is a vertex of $G$, and every edge of $G_H$ is an edge in $G$
also.
$G=(V,E)$ is called \textit{Eulerian} if and only if each vertex
of $G$ has even degree. A graph $G$ is \textit{planar} if it can be drawn
in the plane in such a way that no two edges meet each other except at
a vertex. Any such drawing is called a \textit{planar drawing}. Any
planar drawing of a graph $G$ divides the plane into regions, called
\textit{faces}. One of these faces is unbounded, 
and called the \textit{outer face} or \textit{unbounded face}. 
A \textit{geometric dual graph} [\onlinecite{harary89}] $G_D$ of a connected
planar graph $G$ is a graph $G_D$ with the property that it has a
vertex for each face of $G$ and an edge for each edge touching two
neighboring faces in $G$. 

A \textit{matching} in a graph $G=(V,E)$ is a set of edges $M \subseteq
E$ such that no vertex of $G$ is incident with more than one edge in
$M$. A matching $M$ is \textit{perfect} if every vertex is incident to an
edge in the matching. A maximum matching is a matching of maximum
weight $w(M) = \sum\limits_{e \in M} w(e)$. Solving the perfect matching problem
on general graphs in time bounded by a polynomial in the size of the input remained an
elusive goal for a long time until Edmonds [\onlinecite{edmonds65a,
  edmonds65b}] gave the first polynomial-time algorithm \textemdash\
the blossom algorithm. More details about matching theory can be found
in [\onlinecite{lovasz86}].

%---------------------------------------------------------------------------%
%---------------------------------------------------------------------------%
\section{Review of the Known Algorithmic
  Approaches}\label{sec:algorithms} 

Bieche et al.~[\onlinecite{bieche80}] showed that the problem of
finding a ground state for two-dimensional planar Ising spin glasses
can be transformed to a well known graph theoretical problem
\textemdash\ the minimum-weight perfect matching problem (MWPM) on
general graphs. The method follows the scheme shown in Algorithm
\ref{alg:scheme} in which an optimum matching is used to construct a
spin configuration minimizing the total energy.

\begin{algorithm*}
  \caption{\textsc{Calculate a ground state of a }$K\times L$ \textsc{
      spin glass}\protect\label{alg:scheme}}
	\algsetup{
	linenodelimiter=.
	}
	\begin{algorithmic}[1]
		\REQUIRE \textsc{Planar grid graph } $G_{K,L}$
		\ENSURE \textsc{Spin configuration} $\mathcal{S}$ \textsc{ minimizing the total energy} $\mathcal{H}$
		\STATE \textsc{Construct an appropriate dual graph} $\tilde{G}$
		\STATE \textsc{Calculate a minimum-weight perfect matching} $M$ \textsc{ in} $\tilde{G}$
		\STATE \textsc{Use} $M$ \textsc{ to compute a spin configuration} $\mathcal{S}$ \textsc{ and the 
		        corresponding energy} $\mathcal{H}$
		\RETURN $\mathcal{S}$ \textsc{ and} $\mathcal{H}$
	\end{algorithmic}
\end{algorithm*}

Most commonly used exact methods, like the approaches of Bieche et
al.~ [\onlinecite{bieche80}] and Barahona [\onlinecite{barahona82a,
  barahona82b}], follow this scheme.  In the following, we briefly
summarize these two methods. Afterwards, we present a method following
the construction introduced by Kasteleyn [\onlinecite{kasteleyn63}].
More details can be found in the recent tutorial
[\onlinecite{hartmann08}] on algorithms for computing ground states in
two-dimensional Ising spin glasses.

\subsection*{Review of Exact Methods}

\subsubsection{The Approach of Bieche et al.}
\begin{figure}
\includegraphics[scale=0.25]{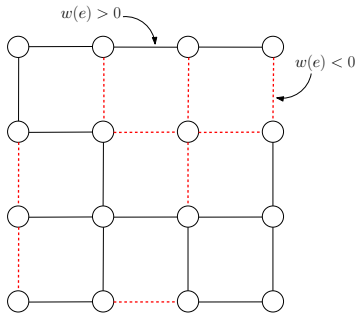}
\caption{$G_{4,4}$ grid graph. Dashed lines indicate negative edge
  weights (Color online). 
  \protect\label{fig:4x4grid}}
\end{figure}

Bieche et al. [\onlinecite{bieche80}] consider the weighted grid graph
$G_{K,L}=(V,E)$ where each vertex $i \in V$ is assigned an initial
spin value $S^0_i=\pm 1$. Each edge $e=(i,j)$ receives a weight
$w(e)=-J_{ij}S^0_iS^0_j$, cf. Fig. \ref{fig:4x4grid}. Often, the
trivial configuration $\mathcal{S}^0 = +1\ \forall S_i,\ i \in V$ is
used.

An instance can not only be described in terms of spins and bonds, but
also by \emph{frustrated plaquettes} and \emph{paths of broken edges}.
Plaquettes consist of the 4-cycles in the
graph.
An edge is said to be satisfied if it attains its minimal weight
($-J_{ij}S^0_iS^0_j = -\lvert J_{ij} \rvert$), otherwise it is called unsatisfied.
A plaquette is frustrated if there is no spin configuration 
satisfying all edges. In this case the plaquette has an odd number
of negative edges.
For the remainder let $F$ be the set of frustrated plaquettes in
$G_{K,L}$ and $P$ the set of all plaquettes in $G_{K,L}$.

\begin{figure}
\begin{minipage}{0.45\textwidth}
  \includegraphics[scale=0.25]{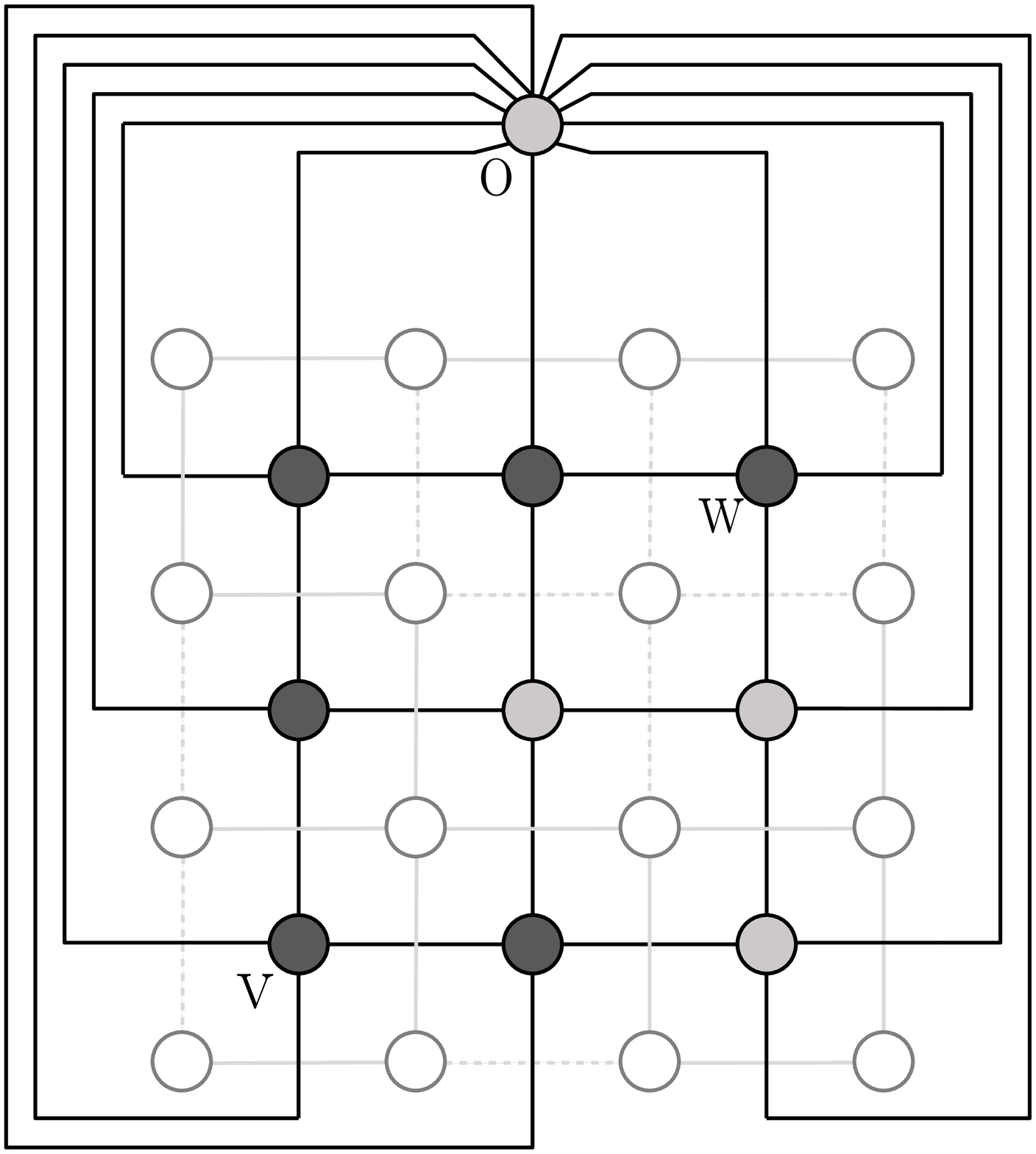}
  \caption{ Geometric dual graph $G_D$ of the grid graph $G_{4,4}$
    shown in Figure \ref{fig:4x4grid} which is seen translucent. Dark
    gray vertices represent frustrated plaquettes (assuming the
    trivial configuration $\mathcal{S}^0 = \{+1\ \mid \ \forall i \in
    V\}$), and light gray vertices are unfrustrated plaquettes
		(Color online).  \protect\label{fig:4x4dual} \vspace{2ex} }
\end{minipage}
\hfill
\begin{minipage}{0.45\textwidth}
 \includegraphics[scale=0.25]{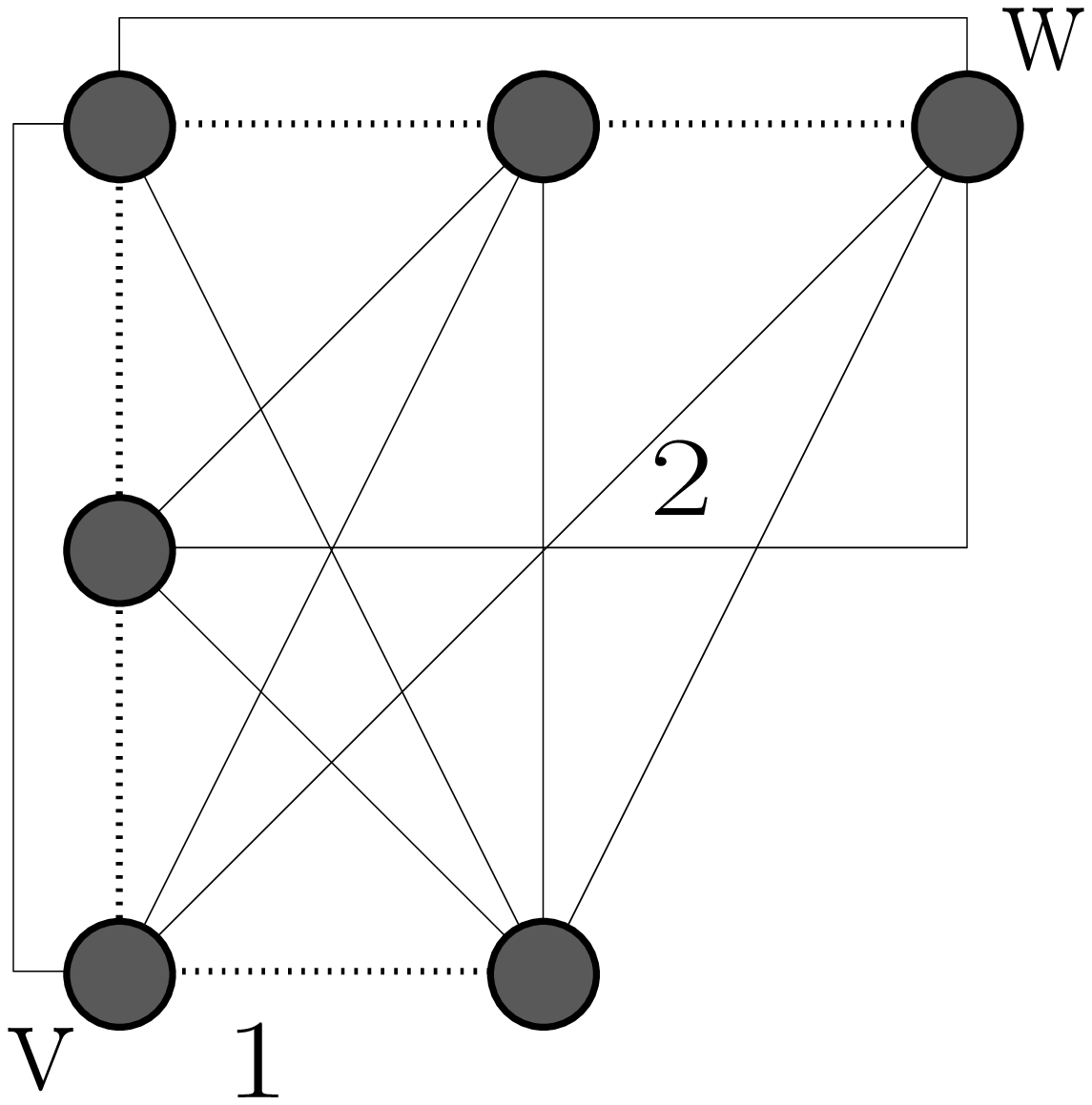}
 \caption{Graph $G_F$ of the grid graph shown in Fig.
   \ref{fig:4x4grid}. Continuous edges indicate distance $2$ between
   vertices, and dotted edges indicate distance $1$ (Color online).
   \protect\label{fig:bieche_dual} }
\end{minipage}
\end{figure}

Bieche et al. identify the frustrated plaquettes as vertices of a 
graph, 
$G_F=(F,E_F)$ with $F=\{f\mid
f\text{ is a frustrated plaquette in }G\}$ and $E_F = F\times
F$. Each edge $e=(f_i,f_j) \in E_F$ is assigned a weight $w(e)$ equal to
the sum of the absolute weights of the edges in $G_{K,L}$ crossed by a
minimum path connecting $f_i$ with $f_j$.
Figure \ref{fig:bieche_dual} shows the graph $G_F$ for the
grid graph of Figure \ref{fig:4x4grid}. The underlying dual graph is shown in Figure 
\ref{fig:4x4dual}.

It is easy to see that minimizing the
sum of the weights of unsatisfied edges connecting frustrated plaquettes 
yields a spin configuration of minimum energy.
The latter is achieved by determining a minimum-weight perfect matching in
$G_F$. Finding a ground state is thus reduced to finding a
minimum-weight perfect matching $M$ of the graph $G_F$, and
its energy is given as:
\begin{eqnarray*}
 \mathcal{H} & = & - \sum\limits_{<i,j>} J_{ij}S_iS_j \\
   & = & - \sum\limits_{<i,j>} \lvert J_{ij}\rvert  + 2 \sum\limits_{\textit{unsatisfied edges}} \lvert J_{ij}\rvert  \\
	 & = & - \sum\limits_{<i,j>} \lvert J_{ij}\rvert  + 2 w(M)
\end{eqnarray*}
For a detailed description of this method we refer to [\onlinecite{hartmann08}].

\subsubsection{Limits of Bieche's Approach}
The approach of Bieche et al. is simple and intuitive, but
comprised two major practical obstacles. First of all, in order to
obtain the dual edge weights, one has to
calculate shortest paths in $G_D$ between all pairs of
vertices. Although this can be done in time and
space bounded by a polynomial in the size of the input, the
calculations can take long in practice or require a large amount of
memory. Equipped with the weights, one can construct the 
complete graph $G_F$ of frustrated plaquettes. 
Treatable system sizes are practically limited
by the number of edges present in $G_F$. Assuming 32 bits for
representing an edge, one needs nearly 4 GB of memory just for
representing $E_F$ for a $300\times 300$ grid graph, assuming
50\% of the plaquettes are frustrated. For a $400\times 400$ grid
graph, almost 12 GB of memory are necessary, which goes beyond the
hardware resources available in ordinary modern computers.

\subsubsection{Simple Improvement for $\pm J$-distributed
  Samples \label{subsec:improvment}}
For $\pm J$ distributed instances, one can obtain the length of the
shortest paths directly without shortest paths calculations.
For this, we project the geometric dual graph $G_D$ on the plane so
that each vertex $v$ is provided with definite coordinates $(x_v,y_v)$
preserving the distance function using rectilinear edges. Different
vertices are assigned to different coordinates.
Then the length of paths between $i$ and $j$, $\pi = (x_i,y_i),
\sigma_1, \sigma_2 \ldots, \sigma_r, (x_j,y_j)$ (with $\sigma_l = (x_l,y_l) = l \in V$)
traversing only through the grid graph $G_{K,L}$ without
crossing its border, is given as the Manhattan distance $c =
\lvert x_i - x_j\rvert  + \lvert y_i - y_j\rvert $. This value is to be compared with the
length of the path passing through the outer face $o$, $\pi_o =
(x_i,y_i), \sigma_1, \ldots, \sigma_k, o, \sigma_{k+1}, \ldots
\sigma_r, (x_j,y_j)$. The weight of this path is given as $c_o =
\min\{x_i, y_i, K-x_i, L-y_i\} + \min\{x_j, y_j, K-x_j, L-y_j\}$. The
shortest path from $i$ to $j$ is the shorter of the two.
In a half-torus graph, analogous calculations can be performed. 

% Consider now a half-torus graph, i.e.~a grid graph with periodic boundary
% conditions in exactly one direction, say the x-axis.
% The shortest path is discovered in the same way, but we are now faced with an additional
% face outside the grid. The weight for the path $\pi$ running only
% through the half-torus graph 
% is given as: $c = \min\{K-\lvert x_i-x_j\rvert, \lvert x_i-x_j \rvert\} + \lvert y_i-y_j\rvert$.
% This has to be compared to the weight $c_o$ of the path $\pi_o$ using the outer face
% vertex with $c_o = L-y_i + L-y_j$ and to the weight $c_{\tilde{o}} = y_i + y_j$ of
% the path using the additional face vertex outside the grid.

\subsubsection{The Approach of Barahona}
Barahona [\onlinecite{barahona82a, barahona82b}] constructs the
geometric dual graph that contains a vertex for each plaquette and edges in
case the corresponding two plaquettes share an edge. Here the outer
face is also interpreted as a plaquette. In formulas, $G_D
= (P,E_D)$ of $G_{K,L}$, where $P = \{p \mid p \text{ is a plaquette
  in } G_{K,L}\} \cup \{o \mid o \text{ is the outer face plaquette}\}$ 
and $E_D = \{e =(p_i,p_j)\ \mid \ \forall p_i,p_j \in
P,\ p_i \cap p_j \neq \emptyset \}$. Each dual edge is assigned a
weight according to the \textit{absolute} weight of the edge in $G_{K,L}$
crossed by the dual edge. Vertices $p_i \in P$ are called \textit{odd} if
they represent a frustrated plaquette, otherwise \textit{even}.

Subsequently, the graph $G_D$ is transformed into a graph $G^{\ast}$.
In order to do this, first every vertex $p_i \in P$ with
$\textit{deg}(p_i) > 3$ is expanded to $(\textit{deg}(p_i)-2)$ copies
of degree $3$. %, cf. Fig.  \ref{fig:barahona_expand}. 
Any even vertex remains even, expanding an odd vertex makes one of its copies
(arbitrarily) odd and the others even. From now on, one works with
vertices of degree 3 only. Next, each vertex is transformed to a $K_3$
subgraph: Each edge incident to an even vertex is replaced by an
intermediate vertex and two edges. At most two new vertices are
inserted for each edge connecting two even vertices. %, cf. Fig.
%\ref{fig:barahona_K3}. 
Original edges keep their weight, new edges
obtain weight zero. For the details, we refer to
[\onlinecite{barahona82a, barahona82b}]. 

%\begin{figure}
%\begin{minipage}{0.45\textwidth}
% \includegraphics[scale=0.25]{barahona_expand.ps}
% \caption{Create for each dual vertex $\textit{deg}(v)-2$ copies
% (Color online). \protect\label{fig:barahona_expand}
% }
%\end{minipage}
%\hfill
%\begin{minipage}{0.45\textwidth}
%  \includegraphics[scale=0.25]{barahona_K_drei.ps}
%  \caption{Expand each vertex (depending on whether it is even or odd)
%  (Color online). \protect\label{fig:barahona_K3}
%  }
% %\end{minipage}
% \end{figure}

On $G^{\ast}$ a MWPM is computed. Any even vertex has an even number (including zero) of
\textquotedblleft outgoing\textquotedblright\ matching edges, however,
any odd vertex has an odd count of those edges. After the matching is
calculated, the afore expanded vertices are shrunken, and the
remaining matching edges raise shortest paths connecting frustrated plaquettes. %, as shown in
%Figures \ref{fig:barahona_matching_1} and \ref{fig:barahona_matching_2}. 
As the total length of the induced paths
is minimal among all possible paths, the induced set of unsatisfied
edges has minimum weight. Following Bieche, this corresponds to a
configuration of minimum weight.

%\begin{figure}
%\begin{minipage}{0.45\textwidth}
% \includegraphics[scale=0.2]{barahona_matching.ps}
% \caption{An odd vertex is expanded to a $K_3$ subgraph, and so each
%   odd vertex has an odd number of incident matching edges
%   (thick lines) (Color online).\protect\label{fig:barahona_matching_1} \\}
%\end{minipage}
%\hfill
%\begin{minipage}{0.45\textwidth}
% \includegraphics[scale=0.2]{barahona_matching_a.ps}
% \caption{Each even vertex is expanded to the described 
%   subgraph. Unlike the odd case, each even vertex has an even number
%   of incident matching edges (thick lines) (Color online).
%   \protect\label{fig:barahona_matching_2} }
%\end{minipage}
%\end{figure}

\subsubsection{Evaluation of Barahona's Approach}
Barahona's transformation consists of two steps and is a bit more
involved than the method of Bieche et al. For a quadratic $L \times L$
grid graph with free boundary conditions, $G^{\ast}$
has $\lvert V^{\ast}\rvert \approx 12(L-1)[(L-1)+2]-12 - b_{odd}$
many vertices, where $b_{odd} = 3\cdot \lvert \{\mbox{odd vertices}\} \rvert$ is the
number of odd vertices in the graph $G_D$. The graph $G^{\ast}$ is sparse as
each vertex in $G^{\ast}$ has degree $3$, $\lvert E^{\ast} \rvert = \frac{3}{2}\lvert V^{\ast}\rvert$.
Assuming $50\%$ frustrated plaquettes, the number of vertices increases approximately 
by a factor of 10.  Given that for bigger lattices this transformation needs less space 
than the one by Bieche et  al., it is preferable to the
former. However, in the next section we describe a method that works with an 
even smaller graph.

\subsubsection{The New Approach \textemdash Following in Kasteleyn's
  Footsteps \label{subsec:kasteleyn}}

In this section, we follow an idea first described by Kasteleyn
[\onlinecite{kasteleyn63, kasteleyn61}] and Fisher
[\onlinecite{fisher66}].  In these works, the goal was to calculate
the configurational partition function for dimer coverings on a
lattice.  The authors exploited that the calculation of the partition
function of the Ising model can be reduced to the number of ways in
which a given number of edges can be selected to form closed polygons
[\onlinecite{domb60}], i.e., a polygon configuration such that each
lattice vertex has even degree of selected incident polygon edges.
The latter can be computed as follows. First, one constructs a
so-called cluster lattice graph which is generated by replacing each
vertex of the lattice by a Kasteleyn city (a $K_4$ subgraph). Now the
expanded graph is oriented such that the associated skew-symmetric
Matrix $D$ shows the property that $\lvert \text{Pf}(D) \rvert$, where
$\text{Pf}(D)$ denotes the Pfaffian of the skew-symmetric matrix,
gives the number of dimer coverings. As there is a one-to-one
correspondence between the number of polygon configurations and the
number of dimer coverings, this method yields the generating function
for polygon configurations and therefore the generating function for
the Ising model.

Thomas and Middleton used Kasteleyn cities for calculating extended ground
states on the torus in order to gain relevant information about
the physics of spin glasses on toroidal lattices. Furthermore, they
point out that the method yields an exact ground-state algorithm on
planar lattices.

A closely related approach was used later by Galluccio et al. to design an
exact algorithm for the computation of the partition function for the
Ising problem that
runs in polynomial time for several models of interest
[\onlinecite{galluccio00, galluccio99I, galluccio99II}], e.g., for
two-dimensional toroidal lattices with $\pm J$ distribution.

\begin{figure}
\includegraphics[scale=0.3]{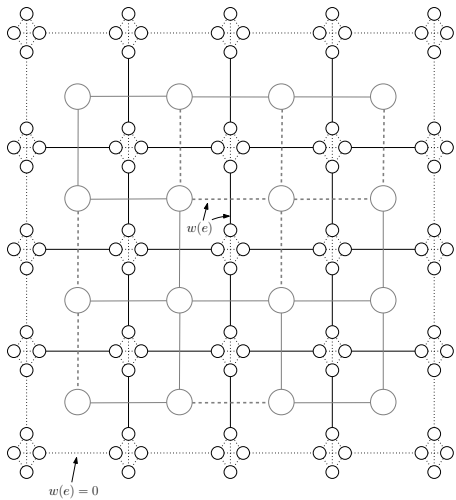}
\caption{Expanded dual graph $G_{K_4}$ for the grid graph $G_{4,4}$
 (Color online). \protect\label{fig:kasteleyn_grid}}
\end{figure}

Here, we focus on planar grid graphs. The distribution of the edge
weights is arbitrary. $G_{K,L}$ is transformed into a
pseudo-dual graph $G_{K_4}$ as shown in Figure \ref{fig:kasteleyn_grid}.
For simplicity, we confine ourselves to quadratic grid graphs in the
following, however all results can be easily transformed to general
grid graphs.

Formally, first the geometric dual of the grid graph $G_{L,L}$ is
constructed, then the outer face vertex is expanded into $2L+1$ ($L$
in the half-torus case) copies, such that the resulting graph is an
intermediate grid graph $G_{L+1,L+1}$ ($G_{L,L+1}$). The edge weights
of $G_{L+1,L+1}$ are set to the weights of the edges of $G_{L,L}$ that
are crossed by the edges of $G_{L+1,L+1}$. Edges that do not cross any
other edge obtain weight zero. Next, each vertex of $G_{L+1,L+1}$ is
expanded to a $K_4$ subgraph. Again, newly constructed edges receive weight zero.

The transformation for the half-torus graph is done similarly, but the
intermediate graph is a \textit{grid-half-torus graph} $G_{L,L+1}$
which is a grid with $L-1$ additional edges. Edge weights are set as just
described. Finally, all vertices are expanded to a $K_4$ subgraph as
before. We denote by $G_{\textit{inter}}$ the intermediate graph
either for the underlying grid or half-torus graph.

On the transformed graph $G_{K_4}$ we calculate a minimum-weight
perfect matching $M$. 

\begin{figure}
\includegraphics[scale=0.4]{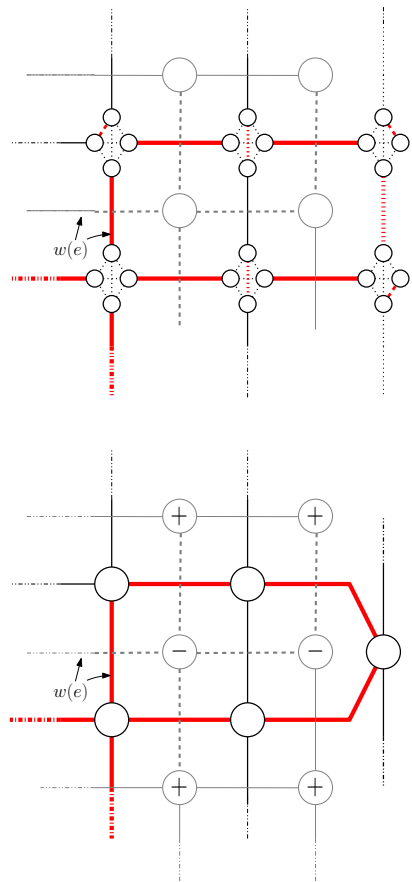}
\caption{Backward transformation. A matching (thick lines) in
  $G_{K_4}$ induces a Eulerian subgraph in $G_D$ and therefore a cut
  in the grid or half-torus graph. The vertex partition consists of
  the vertices with spin value $+1$ ($+$) in one partition and those
  with spin value $-1$ ($-$) in the other (Color online).
  \protect\label{fig:kasteleyn_matching}}
\end{figure}

The next step is to shrink all the $K_4$-subgraphs back, resulting in
the graph $G_{\textit{inter}}$. Also all copies of the outer face
vertex are shrunken. Dealing again with the geometric dual graph of
$G_{L,L}$, we take the subgraph $G_S=(Q,\delta(Q))$ of the geometric
dual graph that consists only of dual edges that were matched, and all
dual vertices with degree greater zero restricted to matched edges. 
This subgraph $G_S$ is an Eulerian graph as each dual vertex is 
incident to an even number of matching edges.
It is well known that there exists 
a one-to-one correspondence between Eulerian subgraphs in the dual
graph and cuts in the original graph. So, $Q$ defines a cut $\delta(Q)$ 
in the graph $G_{L,L}$, cf. Fig. \ref{fig:kasteleyn_matching}.

In order to show the correctness of the transformation, we exploit that
each ground state corresponds exactly to a \textsc{min-cut} $\delta(Q)$ in
the grid graph $G_{L,L}$. 
A cut separates the vertex set into two
disjoint sets $W$ and $V\setminus W$. Vertices in the same partition get
assigned the same spin value. Cut edges are those connecting a pair of
vertices with different spin values. The Hamiltonian can be stated as:
\begin{eqnarray}
\mathcal{H} &=& -\sum_{<i,j>} J_{ij}S_{i}S_{j} \\
 &=& -\sum_{<i,j>} J_{ij} + 2 w(\delta(Q)) \nonumber
\end{eqnarray}

We show that the edge set determined with the method described above
corresponds to a minimum cut. 
\begin{eqnarray}
  w(M)    &=& \sum_{\tilde{e} \in E \cap M} w(\tilde{e}) \nonumber\\
  &\underbrace{=}_{w(\tilde{e}) = w(e)}& \sum_{e \in \delta(Q)} w(e) \nonumber\\
  &=& w(\delta(Q)) \nonumber
\end{eqnarray}

As $w(M)$ is the weight of a minimum-weight perfect matching, the
weight of the subgraph $G_S$ is minimum, and thus also the weight of
the cut $\delta(Q)$.

As Kasteleyn's original method yields the complete
partition function, one might ask why the algorithm was modified so
that only the ground state is determined.
The reason for this is twofold. First of all, 
minimum-weight perfect matchings can be computed in graphs with
millions of vertices, provided they are sparse, which allows
us to go to very large system sizes.
Furthermore, the partition function does not encode the ground-state
configuration itself but only its energy.

\subsubsection{Advantages of the New Approach}
The method is intuitive and its implementation is straightforward. We
present some computational results in Section \ref{sec:compResults}.
For a quadratic $L_1 \times L_1 $ grid graph $G_{L_1,L_1}$ we construct a
graph $G_{K_4}^{L_1,L_1}$ with $\lvert V_{K_4}^{L_1,L_1} \rvert = 4(L_1+1)(L_1+1)$ 
vertices and $2\lvert V_{K_4}^{L_1,L_1} \rvert - 4(L_1+1)$ edges. 
For an $L_2 \times L_2$ ring graph $G_{L_2,L_2}$ the construction
yields a graph $G_{K_4}^{L_2,L_2}$ with 
$\lvert V_{K_4}^{L_2,L_2} \rvert = 4L_2(L_2+1)$ vertices and
$2\lvert V_{K_4}^{L_2,L_2} \rvert - 2L_2 - 4$ edges.
In any case, the resulting graphs are very sparse.
More specifically, let us compare for an $L\times L$ grid the sizes of
the graphs in which a minimum-weight perfect matching is computed.
Assuming that around $50\%$ of the original plaquettes are frustrated,
the graph based on Kasteleyn cities contains only about one
third of the number of vertices and one fourth of the number of
edges contained in the graph constructed with Barahona's method.

As the running time of the matching algorithm scales with the number
of vertices and edges in the graph, the Kasteleyn construction
is preferable to both the Bieche et al.~and the Barahona construction.
 
\subsection*{Computing Domain Walls} 
For the computation of domain walls, we follow the usual approach
[\onlinecite{mcmillan84, riegeretaljuenger96,kawarieger97}].
A ground state of the system is calculated, having energy $E_0$.
To introduce a domain wall, the system is then usually 
perturbed by flipping all couplings along a row or
a column in the lattice. The ground state 
for this new system is calculated, having energy $E^{pert}_0$.
The domain-wall energy for a given sample is then given by 
$\Delta E = \lvert E^{pert}_0 - E_0\rvert$.
We proceed as described in [\onlinecite{hartyoung01,fisch07}], by first determining
a ground state of an EA spin glass with periodic boundary conditions in
one direction, say along the $x$-axis. Then the signs of $L$ edges in
one column along the $y$-axis are flipped.
The symmetric difference of these ground states yields a domain wall.

With the help of linear programming [\onlinecite{chvatal83,nemhauser99}], 
one does not need to calculate the second ground state from scratch 
but can flip the weight of the specific $L$ edges one by one, 
each followed by a reoptimization step. It is also
possible to flip all signs at the same time and do a global 
reoptimization step.
However, in practice for $\pm 1$ distributed weights it often does not pay off to do the
reoptimization steps, and so we calculate both ground states from
scratch.

\subsection*{Modified Approach of Bieche et al. \textemdash\ Review of
  a Heuristic Method}

As argued above, the original approach of Bieche et al. suffers from 
an extensive memory usage. In order to overcome these
limits, some modifications have been proposed that yield heuristic
methods for low-energy states that are however not necessarily exact
ground states. In these heuristics, a reduced graph $\tilde{G}_{red}$
is used instead of a complete one.
% in which only edges are
An approach often used is to introduce in $\tilde{G}_{red}$, only those edges
with weight less than or equal to a fixed value $c_{\max}$
(often $c_{\max}=cJ_{\max}$ is chosen, with $c=4,5,6$).
For continuous spin systems, Weigel [\onlinecite{weigel07}] recently
suggested to introduce for each vertex only the $k$ lightest edges.
He used this cutoff-rule successfully for a matching routine embedded
in a genetic algorithm.

The reasoning behind this is that `heavy'-weighted edges are rarely
contained in an optimum solution. This can assumed to be true for,
e.g., $\pm 1$ distributed couplings and $50\%$ negative weights, for
which very often true ground states are reached in practice. 

Using the reduced graph $\tilde{G}_{red}$, Hartmann and Young
determined high-quality heuristic ground states for $L\times L$
lattices with $\pm 1$ distribution. They could go up to $L\leq 480$
[\onlinecite{hartyoung01}]. Also using the heuristic variant based on
the approach of Bieche et al., Palmer and Adler report results for
$L\leq 1801$ with the choice of $c_{\max}=6J_{\max}$
[\onlinecite{palmeradler99}]. 

For different, especially smaller, percentage of negative weights, the
quality of the heuristic decreases. This can be understood as follows.
An edge $(u,v)$ in the transformed graph is assigned a weight that
equals the sum of the absolute weights of the edges in $G_{K,L}$
crossed by a minimum path connecting $u$ and $v$. $u$ and $v$
correspond to a pair of frustrated plaquettes. The latter can be
assumed to be spread all over the system. In case of small $p$, the
total number of frustrated plaquettes is small. Therefore, the weight
of a minimum path connecting a pair of them can become large which can
cause a heuristic with a limited value of $c_{\max}$ to fail.

Furthermore, for a different distribution of the couplings e.g.,
Gaussian couplings, this heuristic variant has to be used carefully,
as good values for $c_{\max}$ are not evident. Certainly, removing
heavy-weighted edges will result in reduced graphs, but it is not
clear beforehand which weights should be considered heavy. 
Applying different cut-off rules, e.g. vertex-degree constraints,
might be helpful as they were already used for thinning graphs, but
suitable cut-off values depend still on the underlying distribution of the
couplings.

An experimental evaluation will be given in Section
\ref{sec:compResults}.

In the next section, we describe how the correctness of heuristic
ground states can be verified using linear programming.

\subsubsection{\label{subsec:verify}Checking Whether a Spin
  Configuration Actually Defines a Ground State}
Suppose we have a spin configuration at hand that has been computed by
determining an optimum matching on the graph only containing `light'
edges and we want to test whether it is a correct ground state.

First, we compute an optimum matching on the same reduced graph. In
so-called pricing steps, we determine whether yet neglected edges
exist that need to be taken into account in order to ensure correctness. In
case such an edge is reported by pricing, it is introduced in the
reduced graph. The process is iterated until no edge is returned any
more. Pricing is a general feature in linear
programming and combinatorial optimization. For more details, we refer
to [\onlinecite{chvatal83,cook98,nemhauser99}].

Pricing steps can be performed with Cook and Rohe's state-of-the-art
Blossom IV code [\onlinecite{rohe99}] by implementing small
modifications. We give some computational results in Section
\ref{sec:compResults}.

%---------------------------------------------------------------------------%
%---------------------------------------------------------------------------%

\section{\label{sec:compResults}Computational Results}
The method proposed in Section \ref{subsec:kasteleyn} can be used for any
distribution of the couplings. 
Here, we focus on $\pm 1$ distributed instances.
The concentration $p$ of anti-ferromagnetic bonds was set to $0.5$.
For the computations we used Intel Xeon CPU with $2.3$GHz and AMD Opteron Processor $248$ with
$2.2$GHz, each with less than $16$GB RAM.
The largest instances $L > 1500$ were done on Xeon
processors with $16$GB RAM.
The physics analysis done with our method will be presented in a forthcoming
article [\onlinecite{gregorfraukeflorent08}].

Running times for different sizes of grid graphs ($K=L=100$, $150$,
$250$, $500$, $700$, $1000$, $2000$, and $3000$) together with the
number of computed samples, are shown in Table
\ref{tab:gitter-running}.
For smaller grids with $L\leq 500$, we computed between
$2*10^5$ and $5*10^5$ instances per size. For medium sized lattices
$700\leq L\leq 1500$, we ran several thousand samples for each 
size, whereas for the largest systems of $3000^2$ spins we generated
results for 157 samples. Lattices with $L\leq 500$ can be computed within
less than 2 minutes on average, whereas one ground-state determination
for the biggest size requires on average 24h CPU time on a single
processor. Results of samples for spin glasses with periodic boundary
conditions in one direction are presented in Table
\ref{tab:ring-running}. The running times given in Tables \ref{tab:gitter-running} and
\ref{tab:ring-running} scale with $L^{3.50(5)}$ for $L\times L$ spin
glasses. In total we invested about 600 CPU days for
our experiments. 

It turns out that free boundary samples are computationally a bit more
demanding than samples with periodic boundaries in one direction
because the intermediate graphs $G_{\textit{inter}}$ are larger. For
other concentrations of anti-ferromagnetic bonds the presented data
(running times, memory usage, etc.) are comparable, as our method,
unlike the method of Bieche et al., does not depend on the
concentration of anti-ferromagnetic bonds but only on the grid-graph
size $KL$.

\begingroup
\squeezetable
\begin{table}
\begin{ruledtabular}
\begin{tabular}{|r|rr|r|r|r|}
\hline
$L_{grid}$  & \multicolumn{2}{r|}{average time}&   min-time    &   max-time  & nr. samples \\
\hline
 100  &     0.67 & ($\pm    0.00$)&        0.12   &        3.47 &  50925 \\
 150  &     2.03 & ($\pm    0.01$)&        0.31   &       16.80 &  29750 \\
 250  &     9.70 & ($\pm    0.03$)&        1.04   &      121.18 &  36800 \\
 500  &   109.62 & ($\pm    0.79$)&        5.79   &     1406.24 &  20867 \\
 700  &   323.19 & ($\pm    4.54$)&       18.96   &     5233.04 &   5499 \\
 1000 &  1200.33 & ($\pm   17.60$)&       59.49   &     9717.01 &   3483 \\
 1500 &  5280.29 & ($\pm  111.79$)&      288.58   &    58036.33 &   2330 \\
 2000 & 14524.34 & ($\pm  503.69$)&      701.16   &   117313.32 &    942 \\
 3000 & 61166.70 & ($\pm 4920.19$)&     3017.97   &   316581.15 &    157 \\
\hline
\end{tabular}
\caption{Running times and number of computed samples for different
  spin glass instance sizes with free boundary conditions and $\pm 1$
	distributed couplings ($p=0.5$).
\protect\label{tab:gitter-running}}
\end{ruledtabular}
\end{table}
\endgroup

\begin{table}
\begin{ruledtabular}
\begin{tabular}{|r|rr|r|r|r|}
\hline
$L_{ring}$    & \multicolumn{2}{r|}{average time} &   min-time &   max-time &  nr. samples \\
\hline
 400  &   36.00 & ($\pm  1.19$) &    3.27    &     712.67 &   1400 \\
 500  &   88.31 & ($\pm  2.91$) &    5.65    &    1239.81 &   1900 \\
 700  &  319.38 & ($\pm  3.53$) &   17.46    &   12712.34 &  15847 \\
 1000 & 1129.03 & ($\pm 27.20$) &   49.87    &   18577.58 &   3000 \\
\hline
\end{tabular}
\caption{Running times and number of computed samples for different
  spin glass instance sizes with periodic boundary condition in one
  direction and $\pm 1$ distributed couplings ($p=0.5$).
\protect\label{tab:ring-running}}
\end{ruledtabular}
\end{table}

Table \ref{tab:memory_usage} reflects the average over the maximal
memory usage needed in the ground state calculations.
The memory usage roughly scales with $L^{1.9}$ for Ising spin glasses with free boundary 
conditions and with $L^{1.6}$ for Ising spin glasses with periodic boundaries in one direction.
Both from the CPU time and from the necessary memory we conclude that
the method is very fast. It needs considerably less memory than the
commonly used method of Bieche et al., which in its heuristic variant
allows to treat only smaller system sizes than reported here. 
However, we also note that a good statistics for
system sizes beyond $3000^2$ would currently be hard to reach.

\begin{table}
\begin{ruledtabular}
\begin{tabular}{|r|c||r|c|}
\hline
$L_{grid}$ & $\varnothing$ memory & $L_{ring}$  & $\varnothing$ memory \\ 
\hline
100  & 158.7 MB &      & \\
150  & 158.8 MB &      & \\
250  & 163.4 MB &      & \\
     &          & 400  & 245.6 MB \\
500  & 332.5 MB & 500  & 321.6 MB \\
700  & 572.7 MB & 700  & 544.6 MB \\
1000 & 994.6 MB & 1000 & 993.7 MB \\
1500 & 2.052 GB &      & \\
2000 & 3.568 GB &      & \\
3000 & 7.832 GB &      & \\
\hline
\end{tabular}
\caption{Memory usage for different ($\pm 1$, $p=0.5$) sample sizes. 
\protect\label{tab:memory_usage}}
\end{ruledtabular}
\end{table}

\subsection*{Heuristic Ground States and Their Correction}
In this section, we explore the quality of the heuristic ground-state calculation
using the method of Bieche et al.~for
two-dimensional $\pm J$ Ising spin glasses with free boundary conditions.
Within this we use the verification technique explained
in Section \ref{subsec:verify}.

We consider planar $L\times L$ lattices with $L = 164$ and $\pm 1$ distributed couplings. 
First we study these lattices with a concentration $p = 0.5$ of anti-ferromagnetic 
bonds. It turns out that out of $9912$ computed samples $9$
($0.091\%$) were wrong when using $c_{\max}=4$. Thomas and
Middleton [\onlinecite{middleton07}] stated 
$1.5\%$ inexact solutions on toric samples with $L \leq 128$ and
$c=8J_{\max}$. We conclude that in practice the
heuristic almost always returns true ground states if 
$c_{\max}$ and $p$ are suitably chosen.

The overall average running
time was $45.26$ sec., comparing an average of $82.97$ sec. when
pricing was necessary with $45.23$ sec. without pricing.
In our tests, one pricing step was always sufficient to correct a
wrong ground state. Using the Kasteleyn approach, a ground state
computation takes on average only around 1 second 
for this lattice size (on Xeon processors), as can be seen in Table
\ref{tab:kasteleyn-ref}.  

In order to assess the influence of the cut-off parameter $c_{\max}$
on the number of wrong results and the time to correct them, we varied
$p$ and $c_{\max}$ for the $\pm 1\ 164\times 164$ lattices (using the
AMD Opteron processors). In Table \ref{tab:cmax} we show results,
always averaged over $100$ instances.

\begingroup
\squeezetable
\begin{table}
\begin{ruledtabular}
\begin{tabular}{|>{\raggedleft\arraybackslash}p{1.4cm}|rr|>{\raggedleft\arraybackslash}p{1.4cm}|>{\raggedleft\arraybackslash}p{1.6cm}|>{\raggedleft\arraybackslash}p{1.3cm}|} 
\hline
\% edges w. $w(e)=-1$  & \multicolumn{2}{r|}{average time [sec]} & pricing time [sec] &
avg. nr. of pricing steps & nr. of wrong results\\
\hline
  \multicolumn{6}{|c|}{$c_{\max}=3$}  \\
\hline
10 & 16.91 & ($\pm  0.70$) &  9.10 & 2.03 & 100 \\
20 & 51.59 & ($\pm  1.13$) & 18.30 & 1.11 &  93 \\ 
30 & 56.87 & ($\pm  1.57$) & 22.42 & 1.00 &  49 \\ 
40 & 56.95 & ($\pm  1.59$) & 24.01 & 1.00 &  36 \\
50 & 58.72 & ($\pm  1.69$) & 25.10 & 1.00 &  39 \\
\hline
  \multicolumn{6}{|c|}{$c_{\max}=4$}  \\
\hline
10 & 15.60 & ($\pm  0.35$) &  4.76 & 1.09 &  77 \\
20 & 37.37 & ($\pm  0.74$) & 17.99 & 1.00 &   6 \\
30 & 49.93 & ($\pm  0.67$) &  0.00 & 0.00 &   0 \\ 
40 & 53.74 & ($\pm  0.72$) &  0.00 & 0.00 &   0 \\
50 & 54.02 & ($\pm  0.76$) &  0.00 & 0.00 &   0 \\
\hline
  \multicolumn{6}{|c|}{$c_{\max}=5$}  \\
\hline
10 & 13.04 & ($\pm  0.30$) &  4.64 & 1.00 &  18 \\
20 & 37.96 & ($\pm  0.55$) &  0.00 & 0.00 &   0 \\
30 & 52.41 & ($\pm  0.93$) &  0.00 & 0.00 &   0 \\
40 & 56.79 & ($\pm  1.00$) &  0.00 & 0.00 &   0 \\
50 & 59.30 & ($\pm  1.01$) &  0.00 & 0.00 &   0 \\
\hline
  \multicolumn{6}{|c|}{$c_{\max}=6$}  \\
\hline
10 & 13.11 & ($\pm  0.21$) &  5.04 & 1.00 &   1 \\
20 & 42.52 & ($\pm  0.71$) &  0.00 & 0.00 &   0 \\
30 & 59.05 & ($\pm  1.30$) &  0.00 & 0.00 &   0 \\
40 & 65.14 & ($\pm  1.44$) &  0.00 & 0.00 &   0 \\
50 & 62.26 & ($\pm  1.16$) &  0.00 & 0.00 &   0 \\
\hline
  \multicolumn{6}{|c|}{$c_{\max}=7$}  \\
\hline
10 & 14.06 & ($\pm  0.23$) &  0.00 & 0.00 &   0 \\
20 & 46.70 & ($\pm  0.89$) &  0.00 & 0.00 &   0 \\
30 & 66.18 & ($\pm  1.65$) &  0.00 & 0.00 &   0 \\
40 & 73.18 & ($\pm  1.95$) &  0.00 & 0.00 &   0 \\
50 & 69.69 & ($\pm  1.52$) &  0.00 & 0.00 &   0 \\
\hline
  \multicolumn{6}{|c|}{$c_{\max}=8$}  \\
\hline
10 & 15.01 & ($\pm  0.29$) &  0.00 & 0.00 &   0 \\
20 & 53.10 & ($\pm  1.32$) &  0.00 & 0.00 &   0 \\
30 & 74.54 & ($\pm  2.15$) &  0.00 & 0.00 &   0 \\
40 & 83.20 & ($\pm  2.32$) &  0.00 & 0.00 &   0 \\
50 & 82.06 & ($\pm  2.13$) &  0.00 & 0.00 &   0 \\
\hline
\end{tabular}
\caption{Different $c_{\max}$ values used within the heuristic version of the Bieche et al. approach
for each 100 $\pm 1\ 164 \times 164$ grid graphs with various concentration of anti-ferromagnetic bonds.
\protect\label{tab:cmax}}
\end{ruledtabular}
\end{table}
\endgroup

Several conclusions can be drawn from this experiment. Firstly, small
values of $c_{\max}$ lead to many wrong results. E.g., for
$c_{\max}=3$ up to 100\% of the results were wrong, and the
solutions have to be handled with care. Then, the quality of the
heuristic increases with increasing $c_{\max}$. For large enough
value, the solutions are very often correct and only need a
verification step in order to prove their correctness.

Apart from this trend, the results suggest that
the quality of the results highly depends on the chosen
combination of $c_{\max}$ and $p$. 
Clearly, for small percentage $p$ one has to choose a higher cut-off
value $c_{\max}$ in order to generate high-quality solutions. E.g. for
$c_{\max}=3$, all 100 results were wrong when the percentage of
negative edges was chosen as $p=0.1$, whereas 39 results
were wrong for $p=0.5$. This can also be seen by looking at
different $c_{\max}$ values but fixed $p$ values. For $c_{\max} \leq 6$
and $p=0.1$ always at least one percent is wrong.
This means that especially for smaller values of $p$, one
has to make sure that the cut-off value is chosen big enough. The
reason for this behavior is that the weight of a minimum path
connecting a pair of frustrated plaquettes can become large for small
$p$. Therefore, the minimum-weighted perfect matching might contain
heavy edges.  $c_{\max}$ has to be chosen large enough in order to
ensure that these heavy edges are contained in the reduced graph. As
argued before, the necessity of having to choose high cut-off values
can lead to memory problems as the edge density of the generated
reduced graphs increases. These difficulties can be avoided by using
the approach based on Kasteleyn cities. 

From Table \ref{tab:cmax} we see that pricing takes only negligible
running time. As a conclusion, if the Bieche et al.~algorithm is used
to determine ground-state properties, it is advantageous to do the
pricing steps, too, independent of the size of the reduced graph
$\tilde{G}_{red}$. However, despite the fact that pricing is very
fast, the heuristic together with the verification is still
considerably slower than the method proposed here based on Kasteleyn
cities, cf. Table \ref{tab:kasteleyn-ref}.

\begin{table*}
\begin{ruledtabular}
\begin{tabular}{|c|c|c|c|c|c|}
\hline
\backslashbox[0pt][r]{$\lvert V \rvert$}{\% $(w(e) < 0)$} & 10 & 20 & 30 & 40 & 50 \\
\hline
$164^2$ & 0.92 ($\pm 0.005$) & 1.28 ($\pm 0.005$) & 1.32 ($\pm 0.007$) & 1.17 ($\pm 0.007$) & 1.11 ($\pm 0.009$) \\
\hline
\end{tabular}
\caption{Average running times (in sec.) over 10.000 instances, using the method based on Kasteleyn cities
on $\pm 1\ 164 \times 164$ grids with various concentration of anti-ferromagnetic bonds.
\protect\label{tab:kasteleyn-ref}}
\end{ruledtabular}
\end{table*}

It is also interesting to assess the quality of the heuristic for
different distribution of the couplings, e.g. for Gaussian distributed
couplings. We study grid graphs $G_{L,L}$ with $L=50, 100, 150$.
The grid graph size $L$ was limited due to the fact that the generation
of the graphs $G_F$ takes a long time. This is explainable as 
one is in the need of all-pair shortest path calculations for
the set of frustrated vertices on the dual graph. The latter is a
polynomially solvable problem, however the computations can take
long. In our tests, computing the shortest paths usually takes much
longer than computing the ground states itself. In Table \ref{tab:cmax-gauss}
the running times for the matching on the different reduced graphs can be seen.

As it is not clear beforehand how big the cut-off parameter $c_{\max}$
should be, we let ourselves be guided by the corresponding values used
in instances with bimodal distribution. More specifically, for some
value of $c_{\max}$ (we used $c_{\max} = 5$) we compute the average
percentage of 'light' edges from $G_F$ that are contained in the
reduced graph $\tilde{G}_{red}$ in instances with bimodal
distribution. For $L=50$, we find that in $\pm 1$ distributed
instances with $p=0.5$ on average the $8\%$ lightest edges are used
(this percentage reduces to an average of $2\%$ for $L=100$ and to
$1\%$ for $L=150$).

For Gaussian distributed instances, we build the reduced graphs with
the same percentages of light edges, i.e., $8\%$ ($2\%$ or $1\%$)
light edges for $L=50$ ($L=100$, $L=150$ resp.) Results are shown in
Table \ref{tab:cmax-gauss}. First of all, it turns out that for small
grid graphs many results are wrong. For $L=50$, about $12.5\%$ of the
calculated instances do not find correct ground states, and a higher
cut-off value has to be used. Considering larger grid graphs the
situation looks similar. $47\%$ ($52\%$) instances computed the wrong
ground state for $L=100$ ($L=150$).

For $L=50$, we have to go up to a percentage of $16\%$ lightest edges
(corresponding to $c_{\max} \approx 8$), for $L=100$ to $4\%$
(corresponding to $c_{\max} \approx 7$) and for $L=150$ we have to
take into account the $2.5\%$ lightest edges, corresponding to
$c_{\max} \approx 8$, in order to ensure correctness of the results
for our test data. Again, when comparing the running times for $150^2$
lattices, it is by roughly a factor of 25 faster to use the Kasteleyn
city approach instead of a heuristic Bieche method.

As the performance of the matching routine scales with the graph
sizes, more specifically with the number of vertices and edges, we
study the reduced graphs with respect to these two entities.  Usually,
the graphs for computing the matchings are dense.  This is especially
true for small grid graphs ($L=50$), where $16\%$ of the light edges
are needed.  Increasing the grid graphs leads to a considerable
decrease of needed light edges, and decreasing density.  (The density
of a graph with $n$ vertices is defined as the number of its edges
divided by the number of edges of the complete graph $K_n$.)
Nevertheless, in our experiments the grid graphs with $L=100$
($L=150$) contain on average $|V| = 4901\pm 5$ ($|V| = 11103\pm 8$)
vertices.  Taking $4\%$ ($2.5\%$) of all edges means in absolute
numbers $4.803(9)*10^5$ ($1.541(2)*10^6$, resp.)  edges, which are
large and dense graphs. This has to be compared with the graphs used
within the Kasteleyn approach. Here we have for $L=100$ ($L=150$)
graphs with $|V| = 40804$ ($|V| = 91204$) and $8.120*10^4$
($1.818*10^5$) edges. In fact, the graphs we deal with have more
vertices but are considerably sparser. The Blossom IV routine can
compute matchings in very large graphs, provided their density is low.
This fact is also reflected in the better running times for the
Kasteleyn approach presented in this section. Comparing with the
situation for instances with bimodal distributed edge weights, we see
from Table \ref{tab:cmax} that a value $c_{\max} = 4$ in the case of
$p=0.5$ yields good results. For $L=100$ ($L=150$) we have
$|V|=4901\pm 4$ ($|V|=11114\pm 7$) with $|E|=1.58(7)*10^5$
($|E|=3.66(1)*10^5$), which are about $1.3\%$ ($0.6\%$) edges of the
complete graph.  Thus, the graphs with bimodal distribution can be
chosen sparser than in the Gaussian case (for $c_{\max}=4$). However,
these graphs are usually denser than the graphs used in the Kasteleyn
approach.

\begin{table*}
\begin{ruledtabular}
\begin{tabular}{|c|c|r|r|r|}
\hline
$L$ & $c_{\max}$ ($\pm 1$ case) & $\%$ lightest edges (Gassian case) & $\%$ wrong results  & matching time [sec.]\\ 
\hline
50  &  5 &  8.0 & 12.5 &  0.11 ($\pm$ 0.01)\\
    &  6 & 12.0 &  2.0 &  0.18 ($\pm$ 0.01)\\
    &  7 & 15.0 &  2.0 &  0.23 ($\pm$ 0.01)\\
    &  8 & 16.0 &  0.0 &  0.24 ($\pm$ 0.01)\\
\hline
100 &  5 &  2.0 & 47.0 &  2.68 ($\pm$ 0.20)\\
    &  6 &  3.0 &  2.0 &  7.13 ($\pm$ 0.50)\\
    &  7 &  4.0 &  0.0 & 11.27 ($\pm$ 0.80)\\
    &  8 &  5.0 &  0.0 & 23.20 ($\pm$ 1.60)\\
%    & 10 &  8 &  0 \\
\hline
150 &  5 &  1.0 & 52.0 & 34.60 ($\pm$ 2.21)\\
    &  6 &  1.4 &  7.0 & 49.34 ($\pm$ 2.97)\\
    &  7 &  1.9 &  0.0 & 66.70 ($\pm$ 4.00)\\
    &  8 &  2.5 &  0.0 & 85.18 ($\pm$ 5.14)\\
%    & 11 &  4 &  0 \\
%    & 14 &  8 &  0 \\
\hline
\end{tabular}
\caption{Size of reduced graph $\tilde{G}_{red}$ when using Gaussian distributed
couplings compared to the achieved quality by the heuristic variant of the
Bieche et al.~approach.
\protect\label{tab:cmax-gauss}}
\end{ruledtabular}
\end{table*}

%---------------------------------------------------------------------------%
%---------------------------------------------------------------------------%

\section{Conclusions}
We presented a simple algorithm (Section \ref{subsec:kasteleyn}) based
on Kasteleyn cities. The algorithmic foundations of this method date
back to the work of Kasteleyn [\onlinecite{kasteleyn63}] from the 1960s
in which he computed the complete partition function for the Ising model. Using
this approach, we can compute exact ground states for two-dimensional
planar Ising spin-glass instances. The method is easy to implement,
fast and has only limited memory requirements. According to our
knowledge, the treatable system sizes are considerably bigger than the
ones computed earlier and are always provably exact. Thomas and
Middleton [\onlinecite{middleton07}] used Kasteleyn cities for
studying extended ground states. Furthermore, they state that the
method can also be used for determining exact ground states of planar
graphs.

We evaluated different established exact methods and compared them
with respect to running time and memory requirements. It turned out
that the approach presented here is both considerably faster and needs
less memory than the methods proposed earlier. We showed how
heuristically computed ground states can be verified or corrected fast
using mathematical optimization. However, the method based on
Kasteleyn cities still outperforms this approach. Finally, we
evaluated the solution quality of heuristic variants of the Bieche et
al. approach.

In the future, we will make our program available in public domain via
the Cologne Spin Glass Server that can be found at 
\begin{center}
 \verb=http://cophy.informatik.uni-koeln.de/=.
\end{center}

%---------------------------------------------------------------------------%
%---------------------------------------------------------------------------%

\section*{Acknowledgments}

We thank Michael J\"unger for fruitful discussions and Vera Schmitz
for adapting Blossom IV's pricing method as described above.
Thanks to Frank Baumann and Olivier C.~Martin for commenting
on an earlier version of this article and to Alan Middleton for
helpful communications. We are indebted
to two anonymous referees for their valuable comments. 
Last but not least, we thank Oliver Melchert for stimulating discussions and
for providing us with some ground-state data. 
Financial support from the German Science Foundation is acknowledged under
contract Li~1675/1-1. Partially supported by the Marie
Curie RTN Adonet 504438 funded by the EU.

%\bibliography{calculationOf2DplanarIsingSpinglassGroundStates.bib}
%\bibliographystyle{apsrmp}

\end{document}